\documentclass[grl]{AGUTeX}

\usepackage[dvips]{graphicx}

\authorrunninghead{ZHANG et al}

\titlerunninghead{DEMETER OBSERVATION OF PARTICLE BURST}

\begin{document}

%
%

\title{DEMETER Satellite Observations of Particle Burst Prior to Chile Earthquake}
%

%
%


\authors{Zhenxia Zhang, \altaffilmark{1} Xinqiao Li,\altaffilmark{2} Xuhui Shen,\altaffilmark{3}
 Yuqian Ma,\altaffilmark{2} Huaran Chen,\altaffilmark{4} Xinzhao You\altaffilmark{1} and Yahong Yuan\altaffilmark{4}
 }

\altaffiltext{1}{National Earthquake Infrastructure Service, China
Earthquake Administration, Beijing, China. (zxzhang@neis.gov.cn).}

\altaffiltext{2}{Institute of High Energy Physics, Chinese Academy
of Sciences, Beijing, China.}

\altaffiltext{3}{Institute of Earthquake Science,  China
Earthquake Administration, Beijing, China.}

\altaffiltext{4}{Institute of Geophysics, China Earthquake
Administration, Beijing, China}

%
%

\begin{abstract}
The lithosphere activity during seismogenic or occurrence of one
earthquake may emit electromagnetic wave which propagate to
ionosphere and radiation belt, then induce disturbance of electric
and magnetic field and the precipitation of high energy charged
particles. This paper, based on the data detected by DEMETER
satellite, present the high energy charged particle burst(PB) with
4 to 6 times enhancement over the average value observed about ten
days days before Chile earthquake. The obvious particle burst was
also observed in the northern hemisphere mirror points conjugate
of epicenter and no PB events in different years over the same
epicenter region was found. The energy spectra of the PBs are
different from the one averaged within the first three months in
2010. At the same time, the disturbance of the VLF electric
spectrum in ionosphere over the epicenter detected by the DEMETER
satellite are also observed in the same two orbits. Those
observations from energetic PB and VLF electric spectrum
disturbance demonstrates the coupling relation among the
electromagnetic wave emitted by seismic activity, energetic
particle and electric field in ionosphere. We eliminate the
possible origination of PB including magnetic burst and Solar
activities. Finally we think the PB is likely to be related to
Chile earthquake and can be taken as the precursor of this
earthquake.
\end{abstract}

\begin{article}

\section{Introduction}
The correlation between the phenomena of ionospheric disturbance
and seismic activity have been studying for about twenty years,
involving electric and magnetic field variation in horizontal and
vertical components, fluctuations in ion density, thermal change
and energetic particle burst detected on board the ionospheric
satellite. The electromagnetic pulsation at low frequency was
observed on the Intercosmos-Bulgaria-1300 satellite in the
near-equatorial ionosphere over the epicenter of an earthquake
zone on 21st Juanuary 1982(~\cite{Chmerev1989}).
 By ample
statistical data analysis of GEOS-2 satellite, Parrot et.al worked
out that the correlation coefficient between electromagnetic
signal intensification of extreme low frequency and earthquake
with magnitude larger than 4.8 reaches to 5.4
(~\cite{Parrot1985}). The Intercosmos 19 data was researched and
an anomalous increase in the intensity of low-frequency(0.1-16kHz)
radiowave emissions was observed, with the correlation coefficient
of that more than 0.8(~\cite{Larkina1989}). Other important
electromagnetic phenomena (see the papers by
~\cite{Parrot1989},~\cite{parrot2006},~\cite{parrot2009},~\cite{Nemenc2007},~\cite{Nemec2008},~\cite{
zhangxueming2009}) ever observed by satellites also confirmed the
coupling of ionosphere disturbance and seismic activity.

~\cite{Akhoondzadeh2010} observed that the disturbances of the
electron and ion densities in the vicinity of four large
earthquakes and obtained the agreement of anomalies between
DEMETER and GPS(Global Positioning System). The earlier event on
local plasma density fluctuations prior to earthquake was reported
by ~\cite{Gokhberg1983}, which was based on the detection on board
AE-C and ISIS-2 satellites. There are many others papers which
reported the disturbances of ion density in ionosphere associated
with seismic activity several hours to days before the shock by
satellites detections. ~\cite{Shivalika2010} presented and
discussed an anomalous effects of plasma density disturbances and
electric field perturbations associated with Wenchuan earthquake
with the magnitude 7.9 occurred on May 12th, 2008.

As for the high energy particle, ~\cite{voronov1987} and
~\cite{voronov1989} analyzed the data obtained from the MARIA
experiment and reported the correlation between short-term bursts
of energetic charged particle in near-Earth space and seismic
activity for the first time. Later the further study of high
energy charged particle flux was performed by MARIA-2 magnetic
spectrometer on board the MIR station, ELECTRON instruments on
board INTERCOSMOS-BULGARIA-1300 and METEOR-3 satellites.

Using the new experimental results of MARIA-2, GAMMA-1, ELECTRON
and PET instruments, ~\cite{Aleksandrim2003} observed high energy
charged particle bursts in ionosphere and confirmed the evidence
for the temporal and spatial correlation between particle bursts
and seismic activity in agreement with the previous work working
out from the same experiment instruments. ~\cite{lixq} found that
both the electron flux and energy spectrum have significant change
in the NWC VLF electron precipitation belt in the vicinity of
epicenter of Wenchuan earthquake. They also reported that the
400Hz magnetic field appeared the similar evolvement trend with
the energetic particle in the same longitude range.

A review of recent developments of the correlation study between
ionospheric disturbances and seismic activity illuminates that the
events of high energy charged particle bursts associated with
seismic activity are not so many as that of electromagnetic and
ion density variation.

DEMETER (Detection of Electro-Magnetic Emissions Transmitted from
Earthquake Regions) is devoted to the investigation of the Earth
ionosphere disturbances due to seismic and volcanic activities.
The detailed description of its science payloads can be found in
papers(~\cite{Gussac2006},~\cite{Lasoutte2006},~\cite{Sauvaud2006}).
One of the objectives of payload IDP (Instrument for Particle
Detection) on board DEMETER Satellite is to understand the chain
of physical processes leading to the observations before and after
earthquakes. The main objective of another payload ICE experiment
on board DEMETER Satellite is to detect and characterize the
electromagnetic perturbations in the ionosphere that are
associated with seismic activity. Chile earthquake, one of the ten
largest earthquakes in recorded history, had magnitude 8.8 on the
Richter Scale and occurred in the morning 3:34 o'clock on February
27th, 2010 with latitude of $36.1^{\circ} S$ and longitude of
$72.9^{\circ}W$. The corresponding Mc Ilwain parameter value of
epicenter is 1.32 in southern hemisphere.
 In this paper, we present the results in search of PB and VLF
 electric spectrum disturbance which occurred more than ten days
 before Chile earthquake and study the relation between them
 and seismic activity and compare the situation of the
different years.

\section{Observations}
\begin{figure}
\center{
 \noindent\includegraphics[width=20pc]{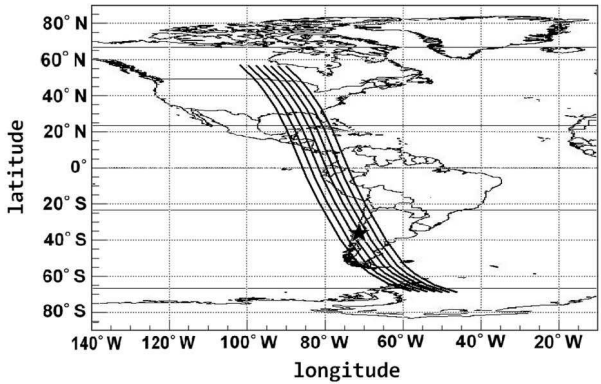}
 \caption{Orbits over epicenter region of Chile earthquake in the first three
 months of 2010. The black pentagram denotes the epicenter position of the quake. \label{orbits}
 }
}
 \end{figure}

\begin{figure}
 \center{
 \noindent\includegraphics[width=15pc]{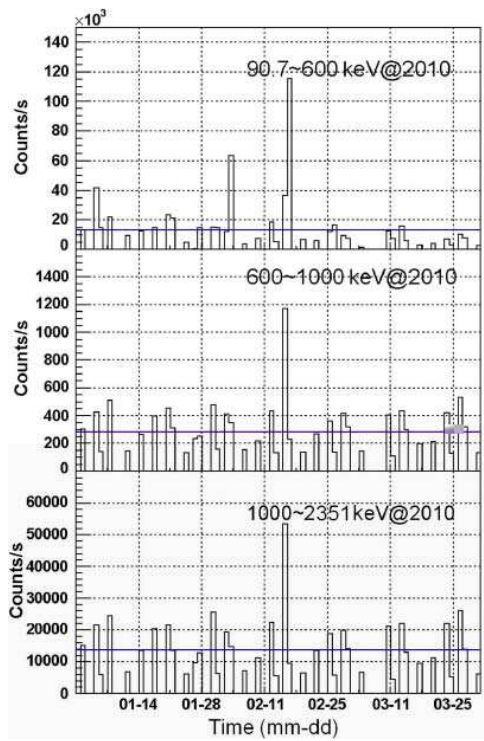}
 \caption{The distribution of high energy charged particle counting rates in 2010
 for three different energy regions. The blue line denotes the
 mean value of counting rates. The peak in the top plot occurred on 16th February from orbit 30109 and
 the peaks in below two plots are occurred on 15th February from orbit 30094. \label{chile-2010}
 }
 }
 \end{figure}

The orbit of DEMETER is quasi sun-synchronous orbit. The half
orbit from northern to southern is downward orbit, and the
corresponding local time is day time. While the corresponding
local time of the half orbit from southern to northern (upward
orbit) is night time. Here we select the satellite detected data
of upward orbit to perform analysis because the VLF electric and
magnetic wave can transmit into the ionosphere in night easier. We
analyzed the high energy particle flux data detected by IDP in the
first three months of 2010. The charged particle counting rates
were considered in three energy regions 90.7$\sim$600keV,
600$\sim$1000keV and 1000$\sim$2351keV, respectively. In order to
obtain the distribution of averaged particle counting rates for
every day, we selected the satellites orbits which fly across the
epicenter region (longitude $72.9^{\circ}W$ and latitude
$36.1^{\circ}S$ or Mc Ilwain parameter 1.32) within the range of
longitude 10 degree and 0.1 Mc Ilwain parameter value. So the
total orbits number of satellite across epicenter of Chile
earthquake during seismic activity in the first three months of
2010 is 42 which are plotted in Figure.~\ref{orbits} . Every
orbits can fly across over the epicenter region of Chile
earthquake for about several minutes.
 From Figure.~\ref{chile-2010} which is the distribution
of high energy charged particle counting rates in 2010, three PBs
are very clearly shown on 16th, February for lower energy region
90.7$\sim$600keV and on 15th, February for two higher energy
regions 600$\sim$1000keV and 1000$\sim$2351keV. The orbit numbers
of three PBs on 15th and 16th February are 30094 and 30109,
respectively. We can found all the three PBs for different energy
region exceed about 4 to 6 times enhancement over the average
value.

\section{The relation of the particle burst and Chile earthquake}
\begin{figure}
\center{
 \noindent\includegraphics[width=15pc]{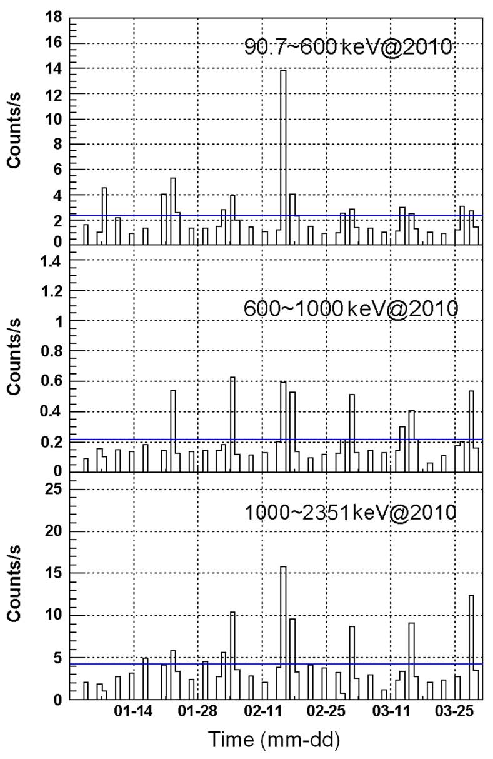}
 \caption{The distribution of high energy charged particle counting rates of the
 northern hemisphere mirror points conjugate of Chile earthquake in 2010
 for three different energy regions. The blue line denotes the
 mean value of counting rates. The peak in the top plot occurred on 14th February from orbit 30079.\label{Chile-mirror-2010}
 }
 }
 \end{figure}

It was known that the charged particles can be trapped by magnetic
fields in a quasi-dipole magnetic field of Earth which conduct
three distinctive motions: gyration, bounce and drift
(~\cite{LiXinLin2001}). Electron drift eastward around the Earth
but proton westward. The trapped particles gyrate around the local
magnetic field when they bounce between the stronger magnetic
fields of Earth in the northern and southern hemispheres. In
northern hemisphere with the same longitude and Mc Ilwain
parameter value to epicenter region, the obvious enhancement of
counting rates of low energy charged particle in 90.7$\sim$600KeV
was discovered 13 days before occurrence of earthquake, shown in
Figure.~\ref{Chile-mirror-2010}, with the significance of more
than 5 times of average value. The corresponding orbit number for
this PB is 30079. There are also obvious energetic particle burst
for the higher energy range of 1000$\sim$2351KeV. It is the bounce
motions that induce the appearance of PB in the mirror region of
epicenter of Chile earthquake.

\begin{figure*}
 \noindent\includegraphics[width=40pc]{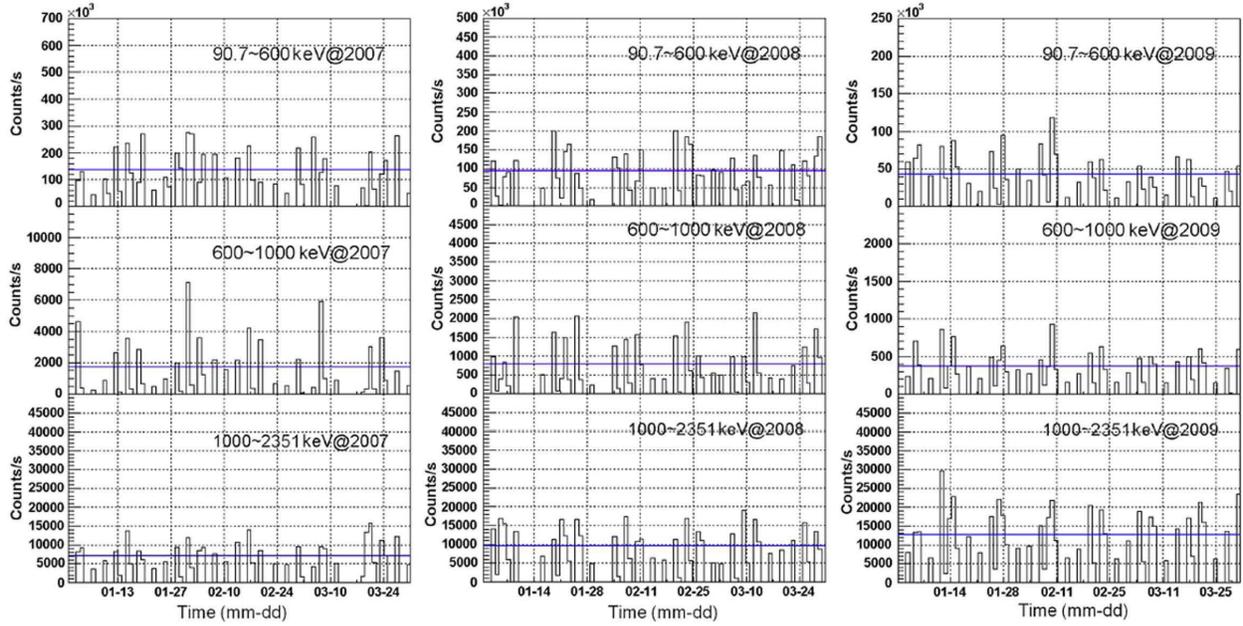}
 \caption{The distribution of high energy charged particle counting rates
 in epicenter region for tree different energy regions. The blue line denotes the
 mean value of counting rates. From left to right plots are the distribution for
 year 2007, 2008 and 2009.\label{signif3}
 }
 \end{figure*}

We also analyzed the IDP data of DEMETER detected in previous
years for the same region and the result were shown in
Figure.~\ref{signif3} which describes the distribution of
energetic particle counting rates in the first three months of the
year 2007, 2008 and 2009, respectively. There are not obvious
enhancement observed except for some irregular statistic
fluctuations.

The magnetic index Kp is officially adopted by the IAGA(the
International Association for Geomagnetism and Astronomy), which
is an usual way to determine if there has been a disturbance in
the Earth's magnetic field and how severe the disturbance is.
Generally the Kp values below 4 indicates little disturbance. We
investigated Kp values and found there no days with Kp larger than
4 during the occurrence of the energetic charged particle burst on
15th and 16th February, 2010. So we can get the conclusion that
the particle burst 11 and 12 days before Chile earthquake is not
caused by disturbance in the Earth's magnetic field. We also
investigated Solar activities which mainly include the energetic
electron flux,  proton flux and Solar flare
during the particle burst prior to Chile earthquake according to
the record of GOES satellite (\textit{http://www.swpc.noaa.gov/}).
There are alos not enough effect from Solar activity to induce the
particle burst reported here.

\section{The energy spectrum characteristic of the particle burst}
\begin{figure}
\center{
 \noindent\includegraphics[width=20pc]{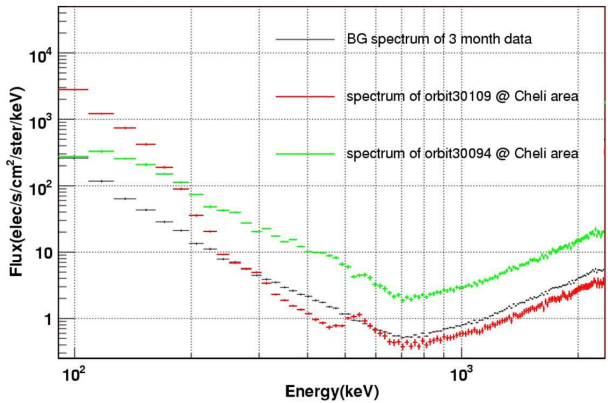}
 \caption{Energy spectrum of high energy charged particle
 detected over epicenter region of Chile earthquake. The black
 profile graph with error bar denotes the average energy spectrum for the first three
 months of 2010 except that of 15th and 16th February, and the red
 one for that of 16th February, the green one for 15th
 February.\label{energyspc}
 }
 }
 \end{figure}

The average energy spectrum of high energy charged particle is
shown in Figure.~\ref{energyspc}. It is detected over epicenter
region of Chile earthquake for the first
 three months in 2010 except the date from orbit of 30109 and 30094 is shown in
Figure.~\ref{energyspc}. Taking this distribution as background,
the energy spectrum change of particle burst on 15th and 16th
February can be easily compared. From Figure.~\ref{energyspc}, we
find that from 15th to 16th February, the energy spectrum in this
region has the evolvement process from soft to hard. The
enhancement of energy spectrum on 15th February is uniform for the
energy larger than 150keV. The enhancement of energy spectrum on
16th February mainly concentrate below the energy of 250keV. This
phenomenon is linked with the coupling relation of wave frequency
and particle energy, which will be discussed later.
 This indicates that the
electromagnetic wave emitted by seismic activity couples firstly
with higher energy particle and then with lower energy particle.


\section{Related observation of VLF electric spectrum disturbance}
 According to the theoretical study of
 ~\cite{gu-vlf}, there is easier coupling between the electromagnetic wave of
  lower frequency and higher energetic particles, while the
  electromagnetic wave of higher frequency has a stronger interaction with lower energetic
  particle. If electromagnetic wave is induced by the seismic activity of Chile earthquake,
   we can deduce that lithosphere in Earth is likely to emit
  lower frequency electromagnetic wave at first, then higher
  frequency of that. The electromagnetic wave emitted by seismic activity may also induce
  the disturbance of electric spectrum. As expected, we surely found the evidence of
  enhancement of VLF electric spectrum E12 corresponding to the energetic
  PB observed in this paper.

\begin{figure}
\center{
 \noindent\includegraphics[width=21pc]{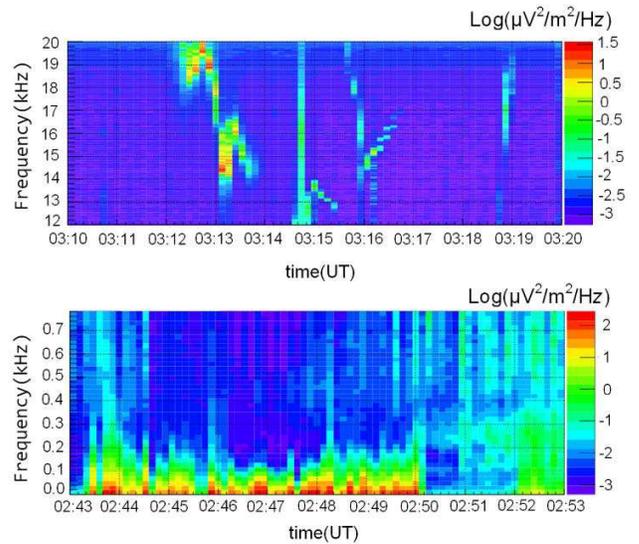}
 \caption{The disturbance of the VLF electric spectrum E12 in ionosphere over
the epicenter detected by the DEMETER satellite. The top plot
comes from orbit number 30109 on 16th February and the below one
from orbit number 30094 on 15th February.
 \label{ice-2orbit}
 }
 }
 \end{figure}

  The observation of the disturbance of the electric field in ionosphere over
the epicenter detected by the DEMETER satellite in the same two
orbit numbers 30109 and 30094 are shown in
figure~\ref{ice-2orbit}. In the top plot related to orbit number
30109 on 15th February, the frequency range with disturbance of
the electric field located at around 14 to 20 kHz and in the below
plot related to orbit number 30094 on 16th February, the frequency
range with disturbance is less than 100Hz. This phenomenon is
consistent with our deduction of the interaction of wave and
particle. What the most important is that there are no similar
disturbance of VLF electric field observed on other orbits flying
across the epicenter position before the earthquake.

\begin{figure}
\center{
 \noindent\includegraphics[width=15pc]{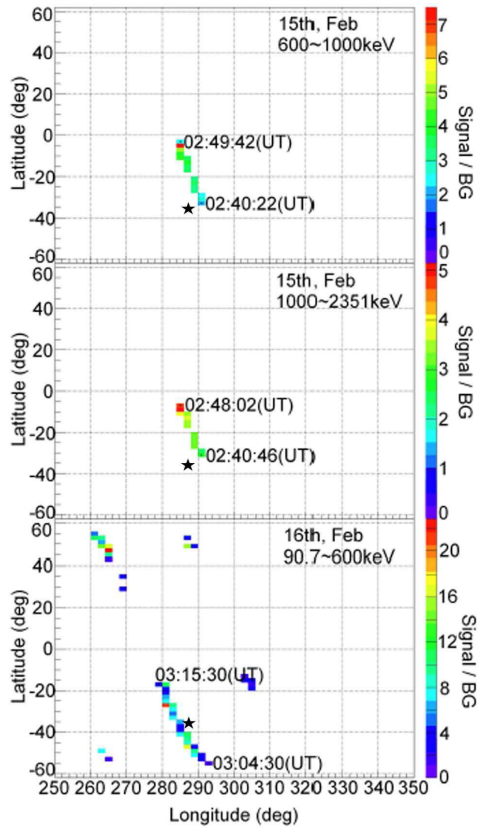}
 \caption{The distribution of position and the duration UT of the energetic particle
 burst in which the value of signal over average background equal or larger than 2 for three energy
 ranges respectively. The black star denotes the position of the epicenter of Chile
 earthquake. The pixel size is $2^{\circ}\times 2^{\circ}$ for latitude and
 longitude.
 \label{pbUT}
 }
 }
 \end{figure}

The particle burst which has the value of signal over local
average background equal or larger than 2 for three energy ranges
respectively are shown in Figure~\ref{pbUT}. The average
background includes the particle counting rates for all local
orbits of the first three months, except the two orbits of 30109
and 30094. The UT time of the appearance of energetic particle
burst in the 90.7$\sim$600keV frequency range is from 03:04:30 to
03:15:30 on 16th February , that in the 600$\sim$1000keV frequency
range from 02:40:22 to 02:49:42 on 15th February, and that in the
1000$\sim$2351keV frequency range from 02:40:46 to 02:48:02 on
15th February. The particle burst for three energy ranges spread
almost in the same longitude region in the center of the
epicenter.

 As for the evolvement of
the disturbance of VLF electric spectrum near the epicenter,
 corresponding to Figure~\ref{ice-2orbit},the
duration UT time of disturbance of the 30094 orbit is from
02:43:42 to 02:50:00 on 15th February. respectively. There are two
parts of VLF electric spectrum enhancement for the 30109 orbit
shown in Figure~\ref{ice-2orbit} and the duration UT time of the
disturbance is from 03:12:00 to 03:13:00 and from 03:13:00 to
03:14:00, respectively.
 We can find that the occurrence of energetic particle burst and the
 disturbance in the VLF electric spectrum is almost simultaneous,
 but on 16th February the energetic particle burst in 30109 orbit appearing a little earlier than
 the VLF electric spectrum disturbance. Because the satellite fly
 from south to north for the upward orbit mode, the later
 appearance of VLF electric spectrum disturbance indicate its
 lower latitude distribution.
 This tells us the phenomenon that the latitude position of
 electric field change is lower than that of the precipitation of energetic particle
 when the electromagnetic wave is transmitted into ionosphere and induce the coupling
 with electric field and energetic particle simultaneously
 ~\cite{sauvaud-NWC} also reported the same
 phenomenon coming from DEMETER satellite data which observed the VLF signal of NWC(US Navy transmitter).
The VLF signal detected by ICE located in the vicinity of south
latitude 17.5 degree and the precipitation of electron in around
south latitude 30 degree. So for the upward orbit at nighttime,
the electron burst signal is detected earlier than VLF electric
signal. This is agreement with the phenomenon reported by this
paper here.

 This indicates that the energetic particle burst and the
 disturbance in the VLF electric spectrum above epicenter of Chile
 earthquake be caused by the same electromagnetic wave which
 was mostly likely to be caused by Chile earthquake, transmitted into ionosphere
 and act with electric field and high energy particle in ionosphere.

\section{Summary}
In this paper, we discovered the energetic charged particle burst
over the epicenter of Chile seismic activity about 11 days before
the quake and the enhancement of them in three energy ranges all
exceed about 4 to 6 times of average value, respectively. In the
mirror points conjugate of epicenter, the particle burst are also
observed obviously with the enhancement in low energy range
90.7$\sim$600keV exceeding 5 times of average value and in high
energy range 1000$\sim$2351keV reaching to around 3 times of
average value. The enhancement of counting rates of the energetic
particle over average background almost occurs in the same
longitude range in the center of the epicenter of Chile
earthquake. After considering every probable production reason
including magnetic burst (magnetic index Kp), Solar activity and
so on, it's likely that this particle burst is caused by Chile
seismic activity and can be looked as the precursor of it.

At the same time, the disturbance of the VLF electric spectrum in
ionosphere over the epicenter detected by the DEMETER satellite
are also observed from the same two orbits. We still studied the
characteristic of energy spectrum of the related orbit with
particle burst on 15th and 16th February and discussed the
coupling relation among the electromagnetic wave emitted by
seismic activity, energetic charged particle and electric field in
ionosphere. We find the phenomenon that the latitude position of
 electric field change is lower than that of the precipitation of energetic particle
 when the electromagnetic wave is transmitted into ionosphere and induce the coupling
 with electric field and energetic particle simultaneously. This
 phenomenon is consistent with the study associated with NWC signal transmitter by
 DEMETER satellite detection.

\begin{acknowledgments}
The authors would like to express their sincere thanks for the
provision of data download of DEMETER Project. This work was
supported by the funding of Science and Technology Supported Plan
of China(2008BAC35B01 and 2008BAC35B05).
\end{acknowledgments}

\end{article}

\end{document}